\documentclass[12pt]{article}
\pdfoutput=1

\usepackage{jheppub}
\usepackage{youngtab}

\title{Line defects and 5d instanton partition functions}

\author[]{Hee-Cheol Kim}

\affiliation[]{Perimeter Institute for Theoretical Physics\\%
31 Caroline Street North, ON N2L 2Y5, Canada}

\emailAdd{heecheol1@gmail.com}

\abstract{We consider certain line defect operators in five-dimensional SUSY gauge theories, whose interaction with the self-dual instantons is described by 1d ADHM-like gauged quantum mechanics constructed by Tong and Wong. The partition function in the presence of these operators is known to be a generating function of BPS Wilson loops in skew symmetric tensor representations of the gauge group. We calculate the partition function and explicitly prove that it is a finite polynomial of the defect mass parameter $x$, which is an essential property of the defect operator and 
the Wilson loop generating function.
The relation between the line defect partition function and the {\it qq}-character defined by N. Nekrasov is briefly discussed.}

\begin{document}
\maketitle

\section{Introduction}

String theory provides a very nice way of realizing solutions to the self-dual Yang-Mills equations, through the low energy dynamics of D$p$/D($p+4$)-brane system. One of the most remarkable results in~\cite{Witten:1995gx,Douglas:1995bn,Douglas:1996uz} is that the space of vacua of the worldvolume theory on a stack of D$p$-brane is shown to be isomorphic to the space of solutions generated by so-called ADHM construction~\cite{ADHM}.

In the work~\cite{Tong:2014cha} by Tong and Wong, the ADHM construction for the self-dual instanton solutions in 5d gauge theories was generalized to include additional charged line defects whose quantization gives rise to BPS Wilson loops in the gauge theories. It was found that the line defects admit a brane realization, originally proposed in~\cite{Gomis:2006sb}, and interactions between the instantonic particles and the defects can be described by one-dimensional $\mathcal{N}=4$ gauged quantum mechanics (QM) on the branes.
In coupling the line defects, the D0-D4 system for the self-dual instantons was modified by adding an extra D4'-brane which intersects with the primary stack of D4-branes at a point on the spatial $\mathbb{R}^4$. New strings stretched between the original D0-D4 system and the new D4'-brane provide extra supermultiplets, as discussed in~\cite{Tong:2014yna} in the T-dual theory, in the standard ADHM QM. In particular, the D4-D4' strings give rise to fermionic degrees of freedom. It was thus argued in~\cite{Gomis:2006sb,Tong:2014cha} that the path integral involving this fermionic Fock space becomes a generating function of half-BPS Wilson loops in anti-symmetric tensor representations. This will be reviewed in Section~\ref{sec:ADHM}.

In this note we will calculate the partition functions of the 5d gauge theories with the 1d line defects on $S^1\times \mathbb{R}^4$. They can be interpreted as Witten indices, or generalized Witten indices with suitable chemical potentials turned on, counting degeneracies of BPS particles interacting with the defects. On the instanton background, the path integral of the 5d theories boils down to the Witten index computation in the modified ADHM quantum mechanics in~\cite{Tong:2014cha}. The Witten indices of supersymmetric quantum mechanics has been extensively studied in~\cite{Hwang:2014uwa,Cordova:2014oxa,Hori:2014tda}. We will use these results to compute the partition functions of our 1d/5d coupled systems.
As discussed, those partition functions (or Witten indices) are expected to be generating functions of BPS Wilson loops in anti-symmetric representations.
The main purpose of this note is to show this property, i.e. (\ref{eq:part-Wilson}), explicitly.

We will verify that the partition function is indeed a polynomial of degree $N$  in the fugacity $x$ characterizing the fermion excitations, which naively seems not to be the case due to superficial singularities of $x$ in its integral expression. Proper treatment of the integration contour, basically instructed by {\it Jeffrey-Kirwan} (JK) residue prescription in~\cite{1993alg.geom..7001J}, removes all possible singularities. We will prove this for 5d $\mathcal{N}=1$ SQCD theories and $\mathcal{N}=1^*$ theories with $U(N)$ gauge group.

We will show that the partition function with the 1d defects obeys a certain functional difference equation which has an intimate connection to a generalization of Baxter T-Q equation determining the spectrum of the quantum integrable system associated to Seiberg-Witten geometry of the 5d gauge theories. It also turns out that the partition function coincides with the 5d version of the {\it qq}-character introduced recently by N. Nekrasov in~\cite{Nekrasov:PI,Nekrasov:2015wsu} and the difference equation can be interpreted as a generalized Dyson-Schwinger equation in the SUSY gauge theory.

\paragraph{Note added} 
While we are writing this paper, we became aware of the related preprints~\cite{Bourgine:2015szm,Nekrasov:2015wsu,Kimura:2015rgi} where the regularity of the {\it qq}-character was explained using different approaches.

\section{ADHM with Wilson Lines}\label{sec:ADHM}
Here we briefly review the ADHM construction of the self-dual $U(N)$ gauge connection in the presence of 1d line defects, i.e. heavy fermionic particles. For more information see \cite{Tong:2014cha}.

Without the defects, an explicit description of the self-dual $k$ instantons in the 5d maximal super Yang-Mills theory is available in the form of a 1d gauged quantum mechanics living on $k$ D0-branes bound to $N$ D4-branes. The quantum mechanics  has $U(k)$ gauge symmetry for the D0-branes and $U(N)$ flavor symmetry for the D4-branes.
Let us consider $N$ D4-branes separated along one of their transverse directions, say $x^9$ direction. This theory has $SO(4)_1\sim SU(2)_1^L\times SU(2)_1^R$ global symmetry rotating the $\mathbb{R}^4$ spatial directions of the D4-branes $x^{1,2,3,4}$, and also $SO(4)_2\sim SU(2)_2^L\times SU(2)_2^R$ global symmetry rotating the four transverse directions to the D4-branes $x^{5,6,7,8}$.
The self-dual instantons in the 5d gauge theory preserve 8 supersymmetries. We write them as $Q_{\dot\alpha}^a$ and $Q_{\dot\alpha}^{\dot{a}}$ where $\dot\alpha,\alpha,\dot{a},a$ denote the indices for the $SU(2)_1^L\times SU(2)_1^R\times SU(2)_2^L\times SU(2)_2^R$ symmetry respectively.
The field content of the quantum mechanics can be read off from the brane configuration.  The fields and their charges are summarized in Table~\ref{tab:field-content} in terms of $\mathcal{N}=4$ supermultiplets. The Higgs branch of this theory is parametrized by the hypermultiplet scalar fields $Z_{\alpha\dot\alpha}$ and $\omega_{\dot\alpha}$, called ADHM data, subject to the D-term and superpotential constraints. It coincides with the moduli space of $k$ instantons.
\begin{table}[!h]
\begin{center}
\begin{tabular}{|c|c|c|c|| c | c | c|c|}
\hline
Multiplet & Field & $U(k)$ & $U(N)$ & Multiplet & Field & $U(k)$ & $U(N)$\\
\hline \hline
Vector & $A_t, \varphi, \lambda_{\dot\alpha}^{\dot{a}} $ & adj & 1 & (Twisted-)Hyper & $Y^{a\dot{a}} , \lambda_{\dot\alpha}^a$ & adj & 1\\
\hline
Hyper & $Z_{\alpha\dot\alpha}, \lambda_\alpha^{\dot{a}}$ & adj & 1  & 
Fermi & $\lambda_\alpha^a$ & adj & 1\\
\hline
Hyper & $\omega_{\dot\alpha}, \psi^{\dot{a}}$ & ${\bf k}$ & ${\bf \bar{N}}$ & Fermi & $\psi^{a}$ & ${\bf k}$ & ${\bf \bar{N}}$\\
\hline
\end{tabular}
\caption{$\mathcal{N}=4$ supermultiplets of $k$ instantons. \label{tab:field-content}}
\end{center}
\end{table}

We now couple the 1d fermionic degrees of freedom to the bulk 5d gauge theory in such a way to preserve half of the supercharges. So it becomes a half-BPS line defect from the 5d field theory point of view. The action for the 1d fermion fields $\chi$ coupled to the 5d bulk fields is given by
\begin{equation}
	S^{1d} = \int dt \, \chi^\dagger (\partial_t - i A_t + \Phi + M)\chi \ ,
\end{equation}
where $A_t$ and $\Phi$ are the pullbacks of the gauge and scalar fields in the 5d vector multiplet. $M$ is the real mass parameter of the fermions, or the background gauge field for the $U(1)$ global symmetry acting only on $\chi$. We focus on the case with $\chi$ in the fundamental representation of the $U(N)$ gauge group.

The coupling a gauge theory to such 1d fermionic degrees of freedom is a classical way to define a BPS Wilson loop. The Fock space of the 1d fermions contains the BPS Wilson loops in representations. Let us insert these 1d fermions into the path integral as
\begin{equation}
	Z^{1d/5d}(M) =  \int \mathcal{D}\Psi \mathcal{D}\chi \ e^{i\left(S_{5d}[\Psi] + S_{1d}[\Psi,\chi,M]\right)} \ ,
\end{equation}
where $\Psi$ stands for the 5d fields.
If we take $L$ excitations of the fermions $\chi$, it inserts a BPS Wilson loop in the $L$-th anti-symmetric representation into the bare partition function~\cite{Gomis:2006sb}. So, the path integral can be schematically written as a polynomial of the fugacity $x\equiv e^M$ counting the number of $\chi$ excitations~\footnote{There could be a global anomaly for the overall $U(1)\subset U(N)$ arising from the contribution of the 1d fermions. Under the large gauge transformation the partition function changes as $Z \rightarrow -Z$, which can be seen from the 1-loop contribution (\ref{eq:1d-contribution}) of the 1d fermions. However, this anomaly is canceled by the induced 1d Chern-Simons term at half-integral level in the our brane system~\cite{Assel:2015oxa}. The author thanks Jaume Gomis for pointing out this issue.}.
\begin{equation}\label{eq:part-Wilson}
	Z^{1d/5d}(x) = x^{-N/2}\sum_{k=0}^N (-x)^k \mathcal{W}_{\Lambda^k} \ ,
\end{equation}
where $\Lambda^k$ denotes the $k$-th anti-symmetric representation and the corresponding Wilson loop is defined as
\begin{equation}
	\mathcal{W}_R = {\rm Tr}_R P\, {\rm exp}\left[i\int dt \, (A_t+i\Phi)\right] \quad {\rm with} \ R=\Lambda^k \ .
\end{equation}
Therefore the partition function of the 1d/5d system gives rise to the generating function of BPS Wilson loops in anti-symmetric tensor representations of the gauge group.

This 1d/5d coupled system admits a brane realization which was first proposed in~\cite{Gomis:2006sb}. In the Type IIA string theory, we consider an additional D4'-brane along $x^{5,6,7,8}$ and time directions. Then the quantization of the string mode stretched between the primary stack of $N$ D4-branes and the other orthogonal D4'-brane will introduce a set of fermionic particles in the worldvolume theory on the D4-branes. They transform as the fundamental representation of the $U(N)$ gauge group. The fermionic degrees of freedom is stuck at the $\mathbb{R}^4$ origin and it has a real mass deformation parameter associated to the relative distance of D4- and D4'-branes along $x^9$-direction. 

When the instantons are coupled to and move around the fermionic degrees of freedom, they feel a Lorentz force proportional to the self-dual gauge connection. For the review of the construction for such self-dual connection and also for the derivation of the low energy effective action on the instanton moduli space in the presence of the fermionic particles, see~\cite{Tong:2014cha} and references therein. We will here review for later computation only the additional field content which are added to the instanton quantum mechanics when coupled to the 1d fermions.

There are now extra fields in the 1d gauged quantum mechanics coming from the strings connecting D0- and D4'-branes and also from the strings between D4- and D4'-branes. These fields are listed in Table~\ref{tab:field-content2}. The fields $\omega',\psi'$ arise from D0-D4' strings and the fields $\chi$ are from D4-D4' strings. The twisted hypermultiplet here means that its matter content is the same as the standard hypermultiplet, but $SU(2)_1^L$ and $SU(2)_2^L$ charges are exchanged. The interaction Lagrangian for these fields was also given in~\cite{Tong:2014yna,Tong:2014cha}.
\begin{table}[!h]
\begin{center}
\begin{tabular}{|c|c|c|c|}
\hline
Multiplet & Field & $U(k)$ & $U(N)$ \\
\hline \hline
(Twisted-)Hyper & $(\omega')^{\dot{a}},\psi'_{\dot\alpha}$ & ${\bf k}$ & 1 \\
\hline
Fermi & $\psi'_\alpha$ & ${\bf k}$ & 1 \\
\hline
Fermi & $\chi$ & 1 & ${\bf N}$ \\
\hline
\end{tabular}
\caption{Additional multiplets due to coupling to the 1d fermions. \label{tab:field-content2}}
\end{center}
\end{table}

\section{Partition functions}
We now consider a supersymmetric partition function of the 5d gauge theory on $S^1\, \times\, \mathbb{R}^4$. We turn on the Omega deformation parameters $\epsilon_1,\epsilon_2$ introduced in~\cite{Nekrasov:2002qd} for the $\mathbb{R}^4$ rotations and also mass parameters $m_a$ for the flavor symmetries.
This partition function can be computed exactly using supersymmetric localization technique. The result was given in~\cite{Nekrasov:2002qd,Nekrasov:2003rj}. 

The partition function after localization takes the form of $Z=Z_{\rm pert}\cdot Z_{\rm inst}$, where $Z_{\rm pert}$ is the perturbative part involving the classical and 1-loop contributions and $Z_{\rm inst}$ is the instanton contribution. The 1-loop contributions for the vector multiplet and fundamental hypermultiplets are given by
\begin{equation}
	Z_{\rm 1-loop}^{\rm vector} = (pq;p,q)_\infty^N \prod_{i\neq j}^N(pq z_i/z_j;p,q)_\infty \ , \quad
	Z_{\rm 1-loop}^{\rm hyper} =\!\! \prod_{i=1}^N\prod_{a=1}^{N_f}(\sqrt{pq}z_i/w_a;p,q)_\infty^{-1} \ ,
\end{equation}
where $(x;p,q)_\infty \equiv \prod_{i,j=0}^\infty(1-xp^iq^j)$. We defined various fugacities as $p=e^{-\epsilon_1},q=e^{-\epsilon_2}, w_a = e^{m_a}$, and $z_i=e^{a_i}$ are the gauge holonomies.

The instanton contribution takes the form of the instanton series expansion as $Z_{\rm inst} = \sum_{k=0}^\infty \mathfrak{q}^k Z_k$ with the fugacity $\mathfrak{q}$ for instanton numbers. $Z_k$ is the $k$ instanton partition function and it can be obtained from the partition function (or Witten index) of the 1d quantum mechanics for $k$ instantons reviewed in the previous section. The partition function of the ADHM quantum mechanics was extensively studied recently in~\cite{Hwang:2014uwa}. We shall review only essential ingredients for our later computation.

It is convenient to first decompose the $\mathcal{N}=4$ multiplets in the ADHM QM into $\mathcal{N}=2$ multiplets and compute contributions from all the $\mathcal{N}=2$ multiplets. Chiral and fermi multiplets in representation $R$ of the $U(k)$ gauge group contribute to the partition function as
\begin{equation}
	Z^{\rm chiral} = \prod_{\rho \in R} 2\sinh\!\left(\tfrac{\rho(\phi) \!+\!2\epsilon_+ J \!+\! 2\epsilon_- \tilde{J}\!+\! m_aF_a}{2}\right)^{-1} \ , \quad
	Z^{\rm fermi} = \prod_{\rho \in R} 2\sinh\!\left(\tfrac{\rho(\phi) \!+\!2\epsilon_+ J \!+\! 2\epsilon_- \tilde{J}\!+\! m_aF_a}{2}\right) \ ,
\end{equation}
where $J$ is the Cartan generator of the diagonal rotation of $SU(2)_1^L\times SU(2)_2^L$ and $\tilde{J}$ is the Cartan generator of $SU(2)_2^R$, and  $F_a$ are the global symmetry generators. $\phi$ is the $U(k)$ gauge holonomy and $\epsilon_\pm \equiv \frac{\epsilon_1\pm\epsilon_2}{2}$. The $\mathcal{N}=2$ vector multiplet contribution is the same as a fermi multiplet contribution. Collecting all contributions we can compute the $k$ instanton partition function with and without the line defect.

In the absence of the BPS line defect, the $k$ instanton partition function takes the form of a contour integral expression~\cite{Nekrasov:2002qd,Kim:2011mv}:
\begin{align}
	Z_k(a,m;\epsilon_{1,2}) &= \frac{1}{k!}\oint \left[\frac{d\phi_I}{2\pi i}\right] \, Z_k^{\rm vector}(\phi,a;\epsilon_{1,2}) \cdot Z_k^{\rm adj}(\phi,a,m;\epsilon_{1,2}) \ ,\cr
	Z^{\rm vec}_k(\phi,a;\epsilon_{1,2}) &= \prod_{I,J=1}^k \frac{\sinh'\frac{\phi_{IJ}}{2}\sinh\frac{\phi_{IJ}+2\epsilon_+}{2}}{\sinh\frac{\phi_{IJ}+\epsilon_1}{2}\sinh\frac{\phi_{IJ}+\epsilon_2}{2}}  \times \prod_{I=1}^k\prod_{i=1}^N\frac{1}{2\sinh\frac{\phi_I-a_i \pm \epsilon_+}{2}} \ , \cr
	Z^{\rm adj}_k(\phi,a,m;\epsilon_{1,2}) &=  \prod_{I,J=1}^k\frac{\sinh\frac{\phi_{IJ}\pm m - \epsilon_-}{2}}{\sinh\frac{\phi_{IJ}\pm m -\epsilon_+}{2}} \times \prod_{I=1}^k\prod_{i=1}^N\sinh\tfrac{\phi_I-a_i \pm m}{2} \ .
\end{align}
Here the prime on the hyperbolic sine indicates that $\sinh(x)$ is omitted when $x=0$. $Z^{\rm vec}_k$ is the contribution from the ADHM fields corresponding to the 5d vector multiplet contribution and $Z^{\rm adj}_k$ is the contribution from the 5d adjoint hypermultiplet with a mass $m$.

This contour integral over the $U(k)$ gauge holonomy $\phi_I$ should be carefully evaluated. It is shown in~\cite{Hwang:2014uwa,Cordova:2014oxa,Hori:2014tda} that the correct contour choice is given by using Jeffrey-Kirwan (JK) prescription first introduced in~\cite{1993alg.geom..7001J} and derived later in~\cite{Benini:2013nda,Benini:2013xpa} for 2d elliptic genera. It turns out that the poles picked up by the JK prescription are classified by so-called $N$-colored Young diagrams. This agrees with the pole prescription in~\cite{Nekrasov:2002qd}. The extra poles provided by the adjoint hypermultiplet contribution yield zero residues and thus only the poles from the standard ADHM fields contribute to the contour integral. See \cite{Hwang:2014uwa} for more detailed explanation.

Combining all the nonzero residues, the $k$ instanton partition for a given $N$-tuple Young diagrams $\vec{Y}=\{Y_1,Y_2,\cdots,Y_N\}$ becomes~\cite{Bruzzo:2002xf}
\begin{equation}
	Z^{\rm inst}_k = \sum_{|\vec{Y}|=k}\prod_{i,j=1}^N\prod_{s\in Y_i}\frac{\sinh\frac{E_{ij}(s)+m-\epsilon_+}{2}\sinh\frac{E_{ij}(s)-m-\epsilon_+}{2}}{\sinh\frac{E_{ij}(s)}{2}\sinh\frac{E_{ij}(s)-2\epsilon_+}{2}} \ ,
\end{equation}
where $|\vec{Y}|$ denotes the total number of boxes in $\vec{Y}$  and
\begin{equation}
	E_{ij}(s)=a_i-a_j-\epsilon_2h_i(s)+\epsilon_1(v_j(s)+1) \ .
\end{equation}
$h_i(s)$ and $v_j(s)$ are the distance from the position $s$ to the right and bottom ends of $i$-th and $j$-th Young diagrams, respectively.

The insertion of the 1d fermions induces additional multiplets into the quantum mechanics listed in Table~\ref{tab:field-content2}. We compute the contribution from these extra multiplets as
\begin{equation}\label{eq:1d-contribution}
	Z^{1d}_k(\phi,a,M;\epsilon_{1,2}) = \prod_{i=1}^N2\sinh\tfrac{a_i-M}{2} \times \prod_{I=1}^k \frac{\sinh\frac{\phi_I-M\pm\epsilon_-}{2}}{\sinh\frac{\phi_I-M\pm\epsilon_+}{2}} \ .
\end{equation}
The full partition function including the contribution from the line defect with mass $M$ is
\begin{align}\label{eq:partition-function-N=1*}
	Z^{1d/5d} &= Z_{\rm 1-loop}^{\rm vector}Z_{\rm 1-loop}^{\rm adj} \cdot Z_{\rm inst}^{1d/5d} \ , \cr
	Z_{\rm inst}^{1d/5d} &= \sum_{k=0}^\infty \mathfrak{q}^k \frac{1}{k!}\oint\left[\frac{d\phi_I}{2\pi i}\right] Z_k^{\rm vector}(\phi,a) \cdot Z_k^{\rm adj}(\phi,a,m) \cdot Z^{1d}_k(\phi,a,M) \ .
\end{align}

The partition function with the 1d fermionic degrees of freedom is a degree $N$ polynomial of the fugacity $x=e^M$, following the discussion in the previous section. The coefficient of $x^L$ in this polynomial is the Wilson loop partition function in the rank $L$ anti-symmetric representation of the $U(N)$ gauge group.
However, one may notice that the line defect contribution (\ref{eq:1d-contribution}) in the contour integral contains the numerator factors which depend on the line defect mass $M$, so the partition function naively becomes an infinite series in $x$ when we Laurent expand it around large $x$. It is thus highly nontrivial to check whether the partition function indeed becomes a finite polynomial in $x$ as required for being physically consistent partition function.
We will prove in the following sections that the partition function is a polynomial of  degree $N$ in $x$ when we take into account the proper contour choice basically determined by the JK prescription.

\section{$\mathcal{N}=1$ $U(N)$ theories}
We first discuss the pure SYM theory with $U(N)$ gauge group. This theory arises from the maximal SYM theory at the energy scale much lower than the adjoint hypermultiplet mass $m$. We will insert the BPS line defects discussed above into this theory and compute the partition function.

We can compute the partition function of the pure SYM theory by taking the limit $m\rightarrow \infty$ of the $\mathcal{N}=1^*$ partition function computed in the previous section. At large $m$, the instanton partition function in (\ref{eq:partition-function-N=1*}) reduces to
\begin{equation}\label{eq:integral-expression}
	Z_{\rm inst}^{N_f=0} =\sum_{k=0}^\infty \mathfrak{q}^k\frac{1}{k!} \oint \left[\frac{d\phi_I}{2\pi i}\right] Z_k^{\rm vector}(\phi,a) \cdot Z_k^{\rm 1d}(\phi,a,M) \ .
\end{equation}
In this limit, as expected, the adjoint hypermultiplet contribution is truncated and we end up with the partition function of the pure SYM theory with the 1d fermions.

Let us now evaluate the contour integral and also prove that the partition function of the 1d/5d coupled system is a finite polynomial in the fugacity $x$. To perform the contour integral over $k$ holonomy variables $\phi_I$, we first need to discuss relevant poles which we should pick up following the JK residue rule. It was observed in~\cite{Hwang:2014uwa} that when we align a reference charge vector $\eta$ for the $U(1)^k\subset U(k)$ gauge group with  Fayet-Iliopoulous (FI) parameter $\zeta$ and choose $\zeta>0$, we should pick up the $k$ poles from chiral multiplets with non-degenerate charge vectors $Q_I$ if they satisfy the following constraint:
\begin{align}
	\eta = (1,1,\cdots,1) = \sum_{I=1}^k a_IQ_I \,, 
\end{align}
where $a_I(>0)$ are certain positive integer numbers. Here the pole and the corresponding charge vector $Q$ are determined by the following equation:
\begin{equation}
	Q(\phi) + \epsilon_+ +\cdots =0 \ ,
\end{equation}
if it comes from a chiral multiplet in a $\mathcal{N}=4$ hypermultiplet, or
\begin{align}
	Q(\phi) - \epsilon_+ +\cdots =0 \ ,
\end{align}
if it comes from a chiral multiplet in a $\mathcal{N}=4$ twisted hypermultiplet.
By summing over all possible poles $\phi_*$
 , we can write the instanton partition function as
\begin{equation}\label{eq:pure-partition-function}
	Z_{\rm inst}^{N_f=0} = \sum_{k=0}^\infty \mathfrak{q}^k\frac{1}{k!} \sum_{\phi_*} \text{JK-Res}_{\phi_*}(Q_*,\eta) \ Z_k^{\rm vector} \cdot Z_k^{\rm 1d} \ ,
\end{equation}
where the JK residue at the given poles is defined as~\cite{1993alg.geom..7001J}
\begin{equation}
\text{JK-Res}_{\phi_*}(\mathrm{Q}_*,\eta)\frac{d^k\phi}{Q_{I_1}(\phi)\cdots Q_{I_k}(\phi)}   = 
\left\{ \begin{array}{cl} \left|{\rm det}(Q_{I_1},\cdots, Q_{I_k})\right|^{-1} & \ \text{if} \ \eta\in \text{Cone}(Q_{I_1},\cdots,Q_{I_k}) \\ 0 & \ \text{otherwise} \end{array}\right.
\end{equation}
`Cone' denotes the cone formed by the $k$ non-degenerate $Q_I$'s.
We also note that there is no subtle wall-crossing issue in our problem since the poles at infinity have zero residue.

For our $U(k)$ gauged quantum mechanics without the extra fields arising from the 1d defect in Table~\ref{tab:field-content2}, the JK prescription reproduces the Young diagram sum rule of the instanton partition function given in~\cite{Nekrasov:2002qd}. One can show that the poles having nonzero JK residue can be specified by the $N$-colored Young diagrams $\vec{Y}=\{Y_1,Y_2,\cdots, Y_N\}$ with total size $k$~\cite{Hwang:2014uwa}. Namely, the $k$ poles satisfy 
\begin{equation}\label{eq:poles-Y}
	\phi_I = a_i + \epsilon_+ - r\epsilon_1-s\epsilon_2 \quad {\rm with} \ (r,s) \in Y_i \ .
\end{equation}

When the extra fields are inserted, their contribution $Z^{\rm 1d}_k$ develops additional poles in the contour integral. The JK pole prescription implies that we should pick up the extra poles obeying the equation
\begin{equation}\label{eq:pole-1d}
	\phi_I - M -\epsilon_+ = 0\ ,
\end{equation}
as well as the poles coming from the original ADHM fields. We find that these extra poles give nonzero contribution to the  partition function. Thus the Young diagram sum rule breaks down, by the JK prescription, when  the 1d defect contribution is inserted. This fact is crucial to make the partition function be a finite polynomial in $x$.

We also notice that we can in fact pick up at most only {\it one pole} of the form (\ref{eq:pole-1d}) among $k$ integral variables $\phi_I$. Namely, once we choose a pole at $\phi_1=M+\epsilon_+$, then the other poles for $k-1$ variables $(\phi_2,\phi_3,\cdots,\phi_{k})$ should be chosen within the Young tableaux classification given in (\ref{eq:poles-Y}) with $|\vec{Y}|=k-1$. One can simply prove this as follows.

We can pick up the first pole at $\phi_1=M+\epsilon_+$ and try to pick up the second pole at $\phi_2-M-\epsilon_+=0$ or $\phi_2 - \phi_1+\epsilon_1=0$, or $\phi_2-\phi_1+\epsilon_2=0$. However, these trial second poles are absent after the first contour integral for $\phi_1$ because they are all canceled by the zeros at $\phi_1-\phi_2=0$ and at $\phi_2-M\pm\epsilon_- = 0$ in the numerator factors.
Therefore, the second and the other poles should be independent of $M$ and $\phi_1$, and should be taken only from the Young diagram with size $k-1$. 
The pole of the form (\ref{eq:pole-1d}) can be chosen at most only once.

Now we are ready to show that the partition function $Z^{1d/5d}$ in (\ref{eq:pure-partition-function}) is a degree $N$ polynomial in the fugacity $x$. The main idea is to first show that the partition function has no pole of $x$ and then study asymptotics of the partition function.
For convenience, we assume that the $k$-th contour includes the poles from $Z_k^{1d}$ as well as the poles in $\vec{Y}$ and the other $k\!-\!1$ contours enclose the poles only in $\vec{Y}$.

The residues at the poles in $\vec{Y}$ can have singularities at
\begin{equation}
	M = a_i - r\epsilon_1-s\epsilon_2 \ , \quad M = a_i -(r-1)\epsilon_1-(s-1)\epsilon_2 \ , \nonumber
\end{equation}
and zeros at
\begin{equation}
	M = a_i - (r-1)\epsilon_1-s\epsilon_2 \ , \quad M = a_i -r\epsilon_1-(s-1)\epsilon_2 \ ,  \quad ({\rm and} \ M=a_i) \ ,\nonumber
\end{equation}
for a given $(r,s)\in Y_i$. A simple algebra can show that most singularities are canceled by the zeros and the remaining poles are located at each convex corner in the $Y_i$. Also all the poles turn out to be non-degenerate.

\begin{figure}[tb]
\centering
\includegraphics[width=6cm]{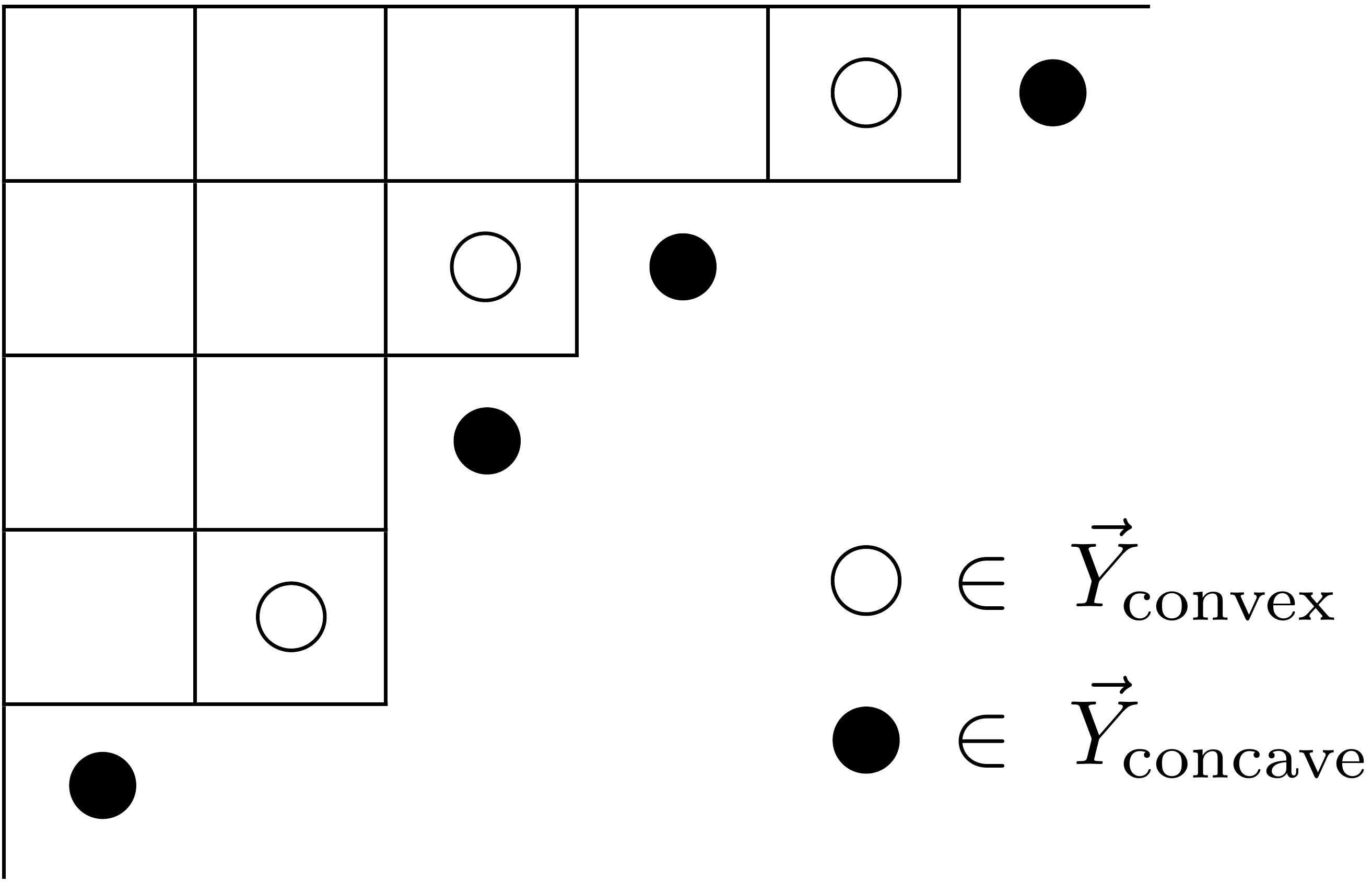}
\caption{ $\vec{Y}_{\rm convex}$ : boxes in convex corners , $ \ \ \vec{Y}_{\rm concave}$ : boxes in concave corners}
\label{fig:young}
\end{figure}

Suppose that the last $k$-th contour integral may lead to a simple pole for $M$ corresponding to $(l,m)$ denoting a box located at one of the convex corners in $\vec{Y}_{\rm convex}$. Any pole after the contour integral can arise when the integration contour is pinched by two poles in the integrand. There are only two sets of poles which can pinch the last contour and develop the pole at $M=a_i-l\epsilon_1-m\epsilon_2$:
\begin{align}
&(1) \ : \qquad \phi_k - M -\epsilon_+ = 0 \ , \quad \phi_k - a_i +\epsilon_+ + l\epsilon_1 + m\epsilon_2  = 0 \cr
&(2) \ : \qquad \phi_k - M +\epsilon_+ = 0 \ , \quad \phi_k - a_i -\epsilon_+ + l\epsilon_1 + m\epsilon_2 = 0 \ .
\end{align}
The first set (1) of poles cannot pinch the integration contour since the second pole in the first set does not exist for the $(l,m)$-th box (located at a convex corner). Furthermore, the second set (2) cannot pinch the contour as well since both poles are inside the integration contour following the JK residue rule explained above. Therefore, the last $k$-th integral cannot produce any new pole for $M$. One may worry about the poles at $(l-1,m)$ and $(l,m-1)$ existing before the last contour integral, but they are always canceled by zeros from the last residue either at $\phi_k\!-\!M\!-\!\epsilon_+=0$ or $\phi_k - a_i + \epsilon_+-l\epsilon_1-m\epsilon_2=0$.
The same argument also holds for all the other possible poles at the convex corners.

From this analysis, we are led to conclude that the partition function evaluated using the JK prescription has no pole in the fugacity $x=e^M$ and thus it is a finite polynomial in $x$. In addition, the asymptotics of the partition function in (\ref{eq:integral-expression}) are
  \begin{equation}\label{eq:asymptotics}
 	x\rightarrow 0 \ : \quad Z_{\rm inst}(x) \rightarrow \mathcal{O}(x^{-N/2}) \ , \qquad
 	x\rightarrow \infty \ : \quad Z_{\rm inst}(x) \rightarrow \mathcal{O}(x^{N/2}) \ .
 \end{equation}
It is therefore obvious that the instanton partition function is a degree $N$ polynomial in $x$ which agrees with the physics in the presence of the 1d fermionic defects. We emphasize that the JK prescription was crucial for this proof. 

The fact that we can choose at most one pole of the form $\phi_I = M+\epsilon_+$ allows us to recast the instanton partition function in an interesting expression. Let us define a new operator as
\begin{align}\label{eq:Y-operator}
	\mathcal{Y}^{\pm1}(M) \ : \ Z_{\rm inst}^{5d} \ \rightarrow \  \sum_{k=|\vec{Y}|=0}^\infty \mathfrak{q}^k \frac{1}{k!} \oint_{\vec{Y}} \left[\frac{d\phi_I}{2\pi i}\right] Z_k^{\rm vector} \cdot \left(Z^{\rm 1d}_k\right)^{\pm1} \ ,
\end{align}
where $\oint_{\vec{Y}}$ means that the integration contours enclose only poles labeled by the given Young diagrams $\vec{Y}$. Note that the 5d partition function after acting $\mathcal{Y}$ operator on differs from the physical partition function $Z^{1d/5d}_{\rm inst}$ by the contour choice. $Z^{1d/5d}$ uses the JK prescription to determine its integration contour, while $\mathcal{Y}$ uses the Young diagram sum rule. 

Using the operator $\mathcal{Y}$, we find that the instanton partition function can be written as
\begin{equation}\label{eq:qq-character}
	Z_{\rm inst}^{N_f=0}(M) = \mathcal{Y}(M) + \mathfrak{q}\frac{1}{\mathcal{Y}(M+2\epsilon_+)} \ ,
\end{equation}
where $\frac{1}{\mathcal{Y}}\equiv \mathcal{Y}^{-1}$. In fact, this coincides with the key characteristic of the five-dimensional {\it qq}-character $\mathcal{X}(x)$ introduced by N. Nekrasov in his recent paper~\cite{Nekrasov:2015wsu}. See also~\cite{Bourgine:2015szm,Kimura:2015rgi} for related discussions.
It turns out that our instanton partition function $Z_{\rm inst}(x)$ with the 1d defect is the same as the {\it qq}-character $\mathcal{X}(x)$, and our operator $\mathcal{Y}$ is identical to his $\mathcal{Y}$-observable~\footnote{In the Nekrasov-Shatashvili limit $\epsilon\rightarrow 0$, the partition functions $Z_{\rm inst}(x)$ and $\mathcal{Y}(x)$ agree with the functions $\chi(x)$ and $\mathcal{Y}(x)$ in~\cite{Nekrasov:2013xda,Nekrasov:2012xe}.}.
More interestingly, the functional difference equation~(\ref{eq:qq-character}) acting on  $\mathcal{Y}$ (by shifting its argument $M$ by $2\epsilon_+$) can be interpreted as the quantization of the defining equation for the Seiberg-Witten curve in the 5d gauge theory~\cite{Nekrasov:2013xda,Nekrasov:2015wsu} with the Planck constant $\epsilon_+$. In the classical limit $\epsilon_+\rightarrow 0$, this relation reduces to the ordinary equation for the Seiberg-Witten curve. This also agrees with the spectral curve of the (relativistic-)closed Toda chain studied in~\cite{Krichever:1999fj}.

We now turn to the $U(N)$ gauge theories with $N_f$ fundamental hypermultiplets. The hypermultiplets provide additional fermionic zero modes in the instanton background.
These fermionic zero modes can be encoded in extra fermi multiplets in the 1d gauged quantum mechanics.
The extra fermi multiplets are given in Table~\ref{tab:Fermi}.
\begin{table}[!h]
\begin{center}
\begin{tabular}{|c|c|c|c|}
\hline
Multiplet & Field & $U(k)$ & $U(N)$ \\
\hline \hline
Fermi & $\xi$ & ${\bf k}$ & 1 \\
\hline
\end{tabular}
\caption{Fermi multiplets induced by 5d fundamental hypermultiplet. \label{tab:Fermi}}
\end{center}
\end{table}
 
The partition function computation with the $N_f$ fundamental hypermultiplets is basically the same as the pure SYM theory cases. We only need to take into account the contribution from the extra fermi multiplet, which can be written as
\begin{equation}\label{eq:partition-hyper}
	 Z_k^{\rm fund}(\phi,m_a;\epsilon_1,\epsilon_2) = \prod_{I=1}^k\prod_{a=1}^{N_f}2\sinh\tfrac{\phi_I-m_a}{2} = \prod_{I=1}^{k} \mathcal{P}(\phi_I) \ , \quad \mathcal{P}(\phi) \equiv \prod_{a=1}^{N_f}2\sinh\tfrac{\phi-m_a}{2} \ .
\end{equation}
We then find that
\begin{equation}
	Z_{\rm inst}^{N_f} =\sum_{k=0}^\infty \mathfrak{q}^k\frac{1}{k!} \oint \left[\frac{d\phi_I}{2\pi i}\right] Z_k^{\rm vector}(\phi,a) \cdot Z_k^{\rm 1d}(\phi,a,M) \cdot Z_k^{\rm fund}(\phi,m_a) \ .
\end{equation}

Since the bulk hypermultiplets induce only fermionic fields in the instanton background, their contribution cannot provide extra singularities to the partition function as one can see from~(\ref{eq:partition-hyper}). This implies that the contour prescription we have discussed for the pure SYM cases still holds for the cases with the additional bulk hypers, unless we have subtle issues related to the small instanton singularity and its regularization~\footnote{The regularization issue about the small $U(1)$ instanton singularity when $N_f\ge 2N - 2|\kappa|$, where $\kappa$ is the Chern-Simons level, has been discussed in literature. See~\cite{Hayashi:2013qwa,Hwang:2014uwa,Gaiotto:2015una} for recent discussions.}. We will discuss only the cases without the subtleties.

Thus, the same argument above for the pure SYM theories proves that the 5d partition function in the presence of the line defect even with the fundamental hypermultiplets is a degree $N$ polynomial of the fugacity $x$. It will give rise to a generating function of BPS Wilson loop expectation values with the hypermultiplets.

The partition function satisfies the relation
\begin{equation}
	Z_{\rm inst}^{N_f}(M) = \mathcal{Y}(M) + \mathfrak{q}\frac{\mathcal{P}(M+\epsilon_+)}{\mathcal{Y}(M+2\epsilon_+)} \ .
\end{equation}
Here the operator $\mathcal{Y}$ defined in (\ref{eq:Y-operator}) inserts the extra factor $Z^{1d}_k$ into the partition function $Z^{N_f}_{\rm inst}$ and  also deforms the integration contour to enclose only the poles labeled by $\vec{Y}$. 


\section{$\mathcal{N}=1^*$ $U(N)$ theories}
We now consider the partition function of the $\mathcal{N}=1^*$ $U(N)$ gauge theories: the $\mathcal{N}=1$ Yang-Mills theory with an adjoint hypermultiplet. 
The full partition function is already given in~(\ref{eq:partition-function-N=1*}).
We shall evaluate the contour integrals and study the properties of the partition function.

We are led  by the JK residue rule to choose poles for the contour integrals of the form
\begin{align}\label{eq:pole-N=1*}
	& (1) \,: \ \phi_I - a_i + \epsilon_+ = 0 \ , \quad (2) \,: \ \phi_I - \phi_J \pm \epsilon_- + \epsilon_+ = 0 \ , \cr
	& (3) \,: \ \phi_I - M -\epsilon_+ = 0 \ , \quad (4) \,:\ \phi_I - \phi_J \pm m - \epsilon_+ = 0 \ .
\end{align}
The first line comes from the standard ADHM fields whereas the second line comes from 
the extra twisted hypermultiplets added by the bulk adjoint hypermultiplet and the 1d line defects. If one chooses all the poles solely in the first line, they can be classified by the $N$-colored Young diagrams $\vec{Y}$ with total size $k$ as in (\ref{eq:poles-Y}). On the other hand, we can also choose the poles only in the second line and then they can be classified by a single Young diagram $\tilde{Y}$ with size $k$, i.e.
\begin{equation}\label{eq:pole-YT}
	\phi_I = M - \epsilon_+ + \tilde{r} (\epsilon_++m) + \tilde{s}(\epsilon_+-m) \quad {\rm with} \ (\tilde{r},\tilde{s})\in \tilde{Y} \ .
\end{equation}
The first case corresponds to all $k$ instantons bound to the stack of $N$ D4-branes, while the second case corresponds to all $k$ instantons bound to a single D4'-brane. 

We will first show that $k$ poles for the $k$ contour integrals should be selected such that when we choose $k'$ of them from a size $k'$ Young diagram $\tilde{Y}$, the other $k-k'$ poles must be chosen from $\vec{Y}$ with size $k\!-\!k'$. There are no other poles having nontrivial residue, and thus the relevant poles for our contour integrals of $U(k)$ holonomies are completely classified by two classes of Young diagrams $\vec{Y}$ and $\tilde{Y}$ with size $k-k'$ and $k'$ respectively. Let us  show this below.

Suppose that the first $k'$ poles are selected only from the $\tilde{Y}$ with size $k'$ and take the form of (\ref{eq:pole-YT}).
Then for the next contour integral, we can choose the poles in the set (4) of (\ref{eq:pole-N=1*}), which increases the size of $\tilde{Y}$ by $+1$, or in the set (1), which starts new Young diagrams $\vec{Y}$, or lastly in the set (2). The pole corresponding to (3) is absent in this case. We want to show that the last cases for the set (2) have zero residues since the poles are canceled by zeros in the integrand. The poles in the set (2) take the following forms
\begin{align}
	&\phi_I - \phi_J \pm \epsilon_- + \epsilon_+ = 0 \cr
	\rightarrow \quad & \left\{
	\begin{array}{c}
	\phi_I -\left(M\!-\!\epsilon_++(\tilde{r}\!-\!1)(\epsilon_+\!+\!m)+\tilde{s}(\epsilon_+\!-\!m) \right) -m \pm \epsilon_- = 0 \\
	\phi_I -\left(M\!-\!\epsilon_++\tilde{r}(\epsilon_+\!+\!m)+(\tilde{s}\!-\!1)(\epsilon_+\!-\!m) \right) +m \pm \epsilon_- = 0
	\end{array}\right.
	\quad {\rm for} \ I>J \ {\rm and} \ (\tilde{r},\tilde{s}) \in \tilde{Y} \nonumber \ .
\end{align}
We note that these poles are always canceled by the zeros in the numerator factors, $\phi_I\!-\!M\!\pm\! \epsilon_-\!=\!0$ and $\phi_I-\phi_{J'} \pm m \pm \epsilon_- \!=\! 0$ with  $ J' \le k'$. Similar argument holds when we first choose $k'$ poles only from $\vec{Y}$ and perform the next contour integral. This shows that the poles are classified by two classes of Young diagrams $\vec{Y}$ and $\tilde{Y}$ with $|\vec{Y}|=k-k'$ and $|\tilde{Y}|=k'$.

Let us now prove that the $\mathcal{N}=1^*$ partition function has no pole in $M$ (or equivalently $x$).
We first consider $k'<k$ integrals evaluated by taking residues at the poles in the class of the $\vec{Y}$ and then attempt to perform the $(k'\!+\!1)$-th contour integral. The discussion for the pure SYM cases in the previous section implies that $(k'\!+\!1)$-th integral will not develop any pole in $M$ independent of the adjoint mass parameter $m$. 
But we may find a new class of poles depending on both $M$ and $m$ which are not excluded by the analysis for the pure SYM cases. 
We need to show that these poles are also absent after computing the remaining integrals over $\phi_{I>k'}$.

Apparently, the first $k'\!+\!1$ integrals cannot yield the new class of poles if we take residues for them at the poles in $\vec{Y}$ since those poles are independent of $M$ and $m$. Also, even if we take the residue for the $(k'\!+\!1)$-th integral at the pole $\phi_{k'+1} = M + \epsilon_+$, it cannot produce the new class of poles since the factors which depend on both $M$ and $m$ (and independent of $\phi_{I>k'+1}$) appear only in the numerator.

For the $(k'+2)$-th integral, one finds that only the following pair of poles can pinch the integration contour $\mathcal{C}$, which is determined by the JK prescription,
\begin{equation}
	\phi_I - M \pm m = 0 \  \not\in \mathcal{C} \ \ , \quad \phi_I -(a_i+\epsilon_+ -l\epsilon_1-n\epsilon_2) = 0 \  \in\mathcal{C} \ ,\nonumber
\end{equation}
 and produce the poles in the new class at $M=a_i\!+\!\epsilon_+\!-\!l\epsilon_1\!-\!n\epsilon_2\pm m$.
The indices $i$ and $(l,n)$ run over all the convex corners in $\vec{Y}_{k'+1}$ $\supset \vec{Y}_{k'}$, where the subscript $k$ in $\vec{Y}_k$ denotes the total size of $\vec{Y}$. 
However, these poles are also canceled by zeros in the numerators of the form
\begin{equation}
	\phi_I - \phi_J \pm m -\epsilon_+ + \epsilon_{1,2} = 0 \quad {\rm at} \quad \phi_{I=k'+1} = M\!+\!\epsilon_+ \ {\rm and} \ \phi_J = a_i\!+\!\epsilon_+ \!-\! r\epsilon_1 \!-\! s\epsilon_2\nonumber
\end{equation}
with $(r,s)\in \vec{Y}_{k'}$. Therefore, the $(k'\!+\!2)$-th contour integral cannot yield any pole in $M$.

The same analysis repeats for $I>k'\!+\!2$ until $I\!=\!k$. The poles in the new class at
\begin{equation}
	M+(\tilde{r}\!-\!1)(\epsilon_+\!+\!m)+(\tilde{s}\!-\!1)(\epsilon_+\!-\!m) = a_i + \epsilon_+-l\epsilon_1-n\epsilon_2\pm m 
	 \ , \nonumber
\end{equation}
for $(\tilde{r},\tilde{s})\in\tilde{Y}$ of size $I\!-\!k'\!-\!1$,
can be generated by the $\phi_{I>k'+2}$ integral.
However, they are all canceled by the zeros in the numerator factors, $\phi_{I-1} \!-\! \phi_J \!\pm\! m \!\pm\!\epsilon_- = 0$, at
\begin{equation}
	  \phi_{I-1} = M\!-\!\epsilon_+\!+\!\tilde{r}(\epsilon_+\!+\!m)\!+\!\tilde{s}(\epsilon_+\!-\!m) \ {\rm and} \ \phi_J = a_i\!+\!\epsilon_+ \!-\!r\epsilon_1\!-\! s\epsilon_2\ \nonumber
\end{equation}
where $(r,s)\in \vec{Y}_{k'}$. So we conclude that the partition function has no pole in $M$. 
Moreover, it follows from two asymptotics at $x\rightarrow \infty$ and $x\rightarrow0$ as shown in (\ref{eq:asymptotics}) that the partition function is indeed a degree $N$ polynomial of the fugacity $x$.

The $\mathcal{N}=1^*$ partition function also satisfies an interesting functional difference equation~\cite{Nekrasov:2015wsu} acting on the operator $\mathcal{Y}$,
\begin{align}
	Z^{\rm inst}(M) = \sum_{\tilde{Y}} \mathfrak{q}^{|\tilde{Y}|} \prod_{s\in\tilde{Y}} \frac{\sinh\frac{E(s)\pm\epsilon_-}{2}}{\sinh\frac{E(s)\pm\epsilon_+}{2}} \times \frac{\prod_{s\in \tilde{Y}_{\rm convex}}\mathcal{Y}(M+F(s))}{\prod_{s\in\tilde{Y}_{\rm concave}}\mathcal{Y}(M+F(s)+2\epsilon_+)} \ ,
\end{align}
where $\mathcal{Y}$ is defined in (\ref{eq:Y-operator}) with the additional contribution from the adjoint hypermultiplet.
$\tilde{Y}_{\rm convex}$ and  $\tilde{Y}_{\rm concave}$ represent two sets of boxes at the convex and concave corners in $\tilde{Y}$, respectively, as depicted in Figure~\ref{fig:young}, and
\begin{equation}
	E(s) = A_+(h(s)+\tfrac{1}{2}) - A_-(v(s)+\tfrac{1}{2}) \ , \quad F(s) = A_+(i-1) + A_-(j-1)
\end{equation}
with $A_\pm\equiv\epsilon_+\pm m$ and $s=(i,j)$.

\section{Conclusion}
In this note we have computed the partition functions of the five-dimensional supersymmetric gauge theories which we can obtain through supersymmetric couplings to 1d fermionic degrees of freedom. We have also proved that the partition function is a finite Laurent polynomial in the fugacity $x$ counting the fermion number. This is an essential requirement for being a physical partition function containing the 1d fermionic Fock space.

One natural question would be whether our approach for BPS line defects generalizes to the gauge theories with other classical gauge groups. The ADHM construction for the instantons in the $SO(N)$ and $Sp(N)$ gauge theories are known in~\cite{Douglas:1996uz,Aharony:1997pm}, and one can naturally couple the additional degrees of freedom in~\cite{Tong:2014cha} as we did for $U(N)$ cases. 
It will be interesting to study properties of the line defect partition functions and their relation with the Seiberg-Witten geometry of $SO(N)$ and $Sp(N)$ theories.

Small instanton singularity requires proper regularization scheme, like ADHM gauged quantum mechanics for $U(N)$ gauge theory without matters and defects. Physically relevant regularization for the instanton singularity has not been well-studied in the presence of BPS operators such as Wilson lines in various representations. It is desirable to understand a general UV prescription for small instantons in the gauge theories with and without (non-)local operators.
Our result provides a natural UV prescription for Wilson lines in anti-symmetric representations which may help to answer that question.

Finally, one can also consider gauge theories that contains multiple defects of different dimensionalities. For example, a 5d guage theory coupled to both the line defects in our note and codimension two defects discussed in~\cite{Gaiotto:2014ina,Bullimore:2014awa} turns out to be supersymmetric. It will be interesting to investigate the partition function of this combined system and its role in the gauge theory and the associated integrable model.

\section*{Acknowledgements}

We would like to thank Jaume Gomis, Seok Kim, Nikita Nekrasov for fruitful discussions and their encouraging comments.
 We are particularly grateful to Davide Gaiotto for collaboration at the early stage of this work and insightful comments.
The research of HK was supported by the Perimeter Institute for Theoretical Physics. Research at Perimeter Institute is supported by the Government of Canada through Industry Canada and by the Province of Ontario through the Ministry of Economic Development and Innovation.

\bibliographystyle{JHEP}

\bibliography{5d-paper}

\end{document}